\documentclass[prl,twocolumn]{revtex4}
\usepackage{graphicx}
\usepackage{amssymb,amsmath,bm}
\usepackage{enumerate}

\newcommand{\be}{\begin{eqnarray}}
\newcommand{\ee}{\end{eqnarray}}

\begin{document}

\title{Turbulence and distributed chaos with spontaneously broken symmetry}

\author{A. Bershadskii}

\affiliation{
ICAR, P.O. Box 31155, Jerusalem 91000, Israel
}

\begin{abstract}

It is shown that in turbulent flows the distributed chaos with spontaneously broken translational  space symmetry (homogeneity) has a stretched exponential spectrum $\exp-(k/k_{\beta})^{\beta }$ with $\beta =1/2$. Good agreement has been established between the theory and the data of direct numerical simulations of isotropic homogeneous turbulence (energy dissipation rate field), of a channel flow (velocity field), of a fully developed boundary layer flow (velocity field), and the experimental data at the plasma edges of different fusion devices (stellarators and tokamaks). An astrophysical application to the large-scale galaxies distribution has been briefly discussed and good agreement with the data of recent Sloan Digital Sky Survey SDSS-III has been established. 
\end{abstract}

\maketitle

"The mathematicians always want that their mathematics should be pure,
 that is, strict 
 and provable, wherever possible.  However, the most interesting and realistic problems could not usually be solved in that manner."  

\~~~~~~~~~~~~~~~~~~~~~~~~~~~~~~~~~~~~~~~~~~~~~~~A.N. Kolmogorov\\

\section{Introduction}

It is known that anisotropic turbulence can emerge from deterministic chaos \cite{swinney1}. In a recent paper Ref. \cite{b1} it was also shown for isotropic homogeneous turbulence. The isotropic homogeneous turbulence emerges from distributed chaos. For the distributed chaos the power spectra  
are linear weighted superposition of the exponentials converging into the stretched exponential:
$$
E(k ) \simeq \int_0^{\infty} \mathcal{P} (\kappa)~ e^{-(k/\kappa)}  d\kappa  \propto \exp-(k/k_{\beta})^{\beta}  \eqno{(1)}
$$
where $\mathcal{P}(\kappa )$ is a probability distribution of $\kappa$, the $\kappa$ is wavenumber of the waves (pulses) driving the chaos. An asymptotic theory has been developed in the Ref. \cite{b1} in order to find $\beta$. In this theory the asymptotic ($\kappa \rightarrow \infty $) scaling of the group velocity of the waves driving the chaos
$$
\upsilon (\kappa ) \sim \kappa^{\alpha}     \eqno{(2)}
$$
is used in order to find the $\beta$:
$$
\beta =\frac{2\alpha}{1+2\alpha}   \eqno{(3)}
$$ 

The main physical problem now is to determine a dimensional parameter controlling the scaling (2).

 \begin{figure}
\begin{center}
\includegraphics[width=8cm \vspace{-1cm}]{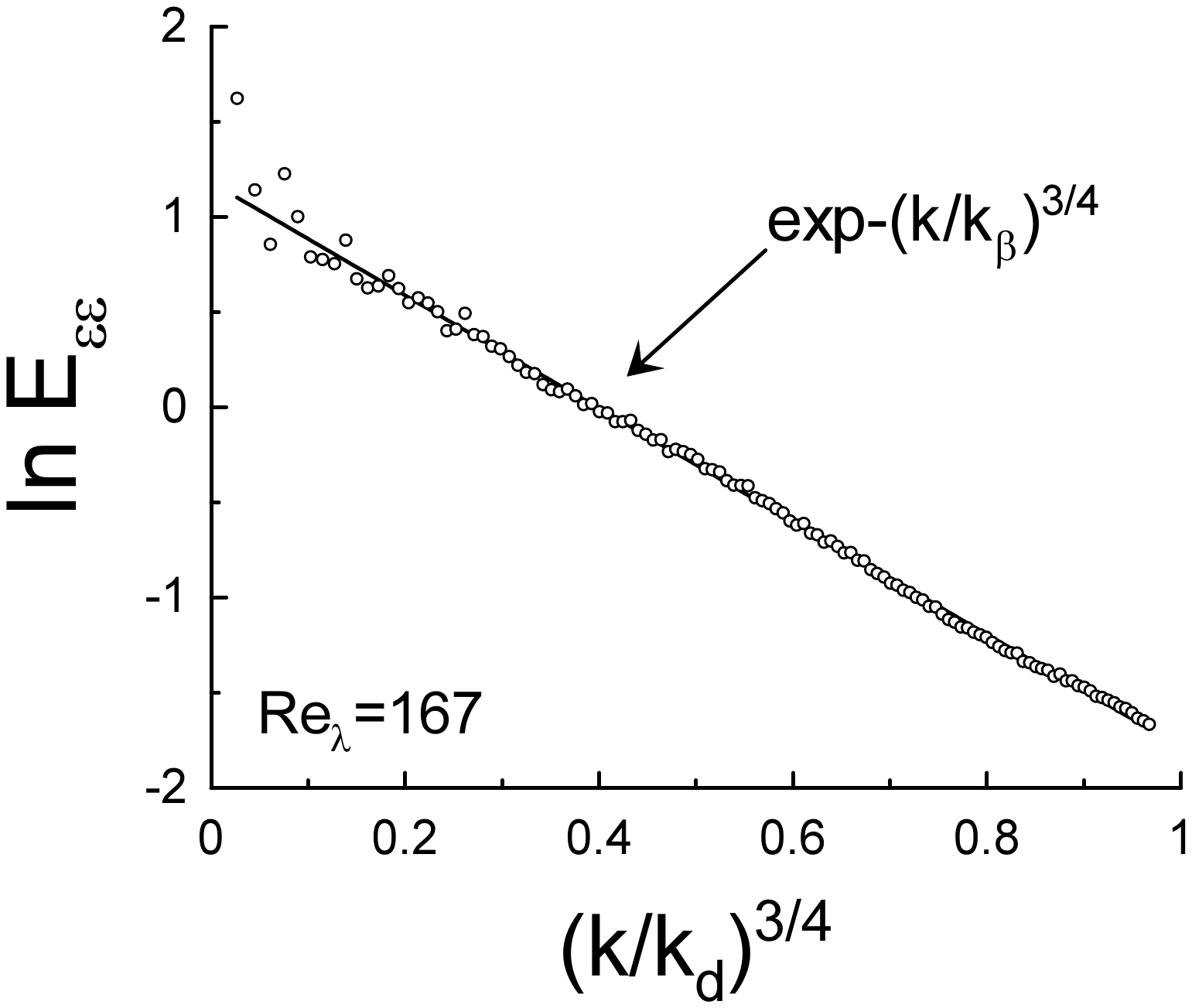}\vspace{-4cm}
\caption{\label{fig1} Logarithm of power spectrum of the energy dissipation rate field $\varepsilon$ at $Re_{\lambda}=167$ as function of $(k/k_d)^{3/4}$. The straight line indicates the exponential decay Eq. (1) with $\beta =3/4$. The data are taken from the DNS \cite{ishi}.}
\end{center}
\end{figure}

 It was recently discovered \cite{nrz} that spontaneous breaking of the space translational symmetry (homogeneity) in wave turbulence results in a crucial change of the statistical attractor. This phenomenon is related to the Kolmogorov-Zakharov spectrum instability to the perturbations weakly breaking the space translational symmetry. Namely, the Fourier transform
$$
\int \langle {\bf u} ({\bf x}) \cdot  {\bf u}^{\star} ({\bf x} + {\bf r}) \rangle e^{i{\bf k}{\bf r}} d{\bf r}
$$ 
of the velocity correlation (where $<...>$ denotes an ensemble average) was allowed to be weakly depending on ${\bf x}$. Spectrum of the state with the spontaneously broken translational symmetry was shown (numerically) to be different from the Kolmogorov-Zakharov spectrum. The authors of the Ref. \cite{nrz} argued that this phenomenon can have a general nature.     

\section{Spontaneously broken space symmetry}

  There are two invariants in isotropic homogeneous turbulence related to the fundamental space symmetries. Let us consider integrals
$$   
I_n = \int r^{n-2} \langle {\bf u} ({\bf x},t) \cdot  {\bf u} ({\bf x} + {\bf r},t) \rangle d{\bf r}  \eqno{(4)}
$$ 
at sufficiently rapid decay of the two-point correlation function of the fluid velocity field. The integrals $I_2$ and $I_4$ are the Birkhoff-Saffman and Loitsyanskii invariants. These invariants are associated with the conservation laws of the linear and angular momentum, respectively \cite{my}-\cite{fl}. These conservation laws themselves (according to Noether's theorem) are consequences of the fundamental space symmetries: translational (homogeneity) and rotational (isotropy) \cite{ll2}. The Birkhof-Saffman invariant determines the scaling Eq. (2) for the isotropic homogeneous turbulence and provides $\alpha = 3/2$ \cite{b1}:
$$
\upsilon (\kappa ) \simeq a~I_2^{1/2}~\kappa^{3/2} \eqno{(5)}
$$
where $a$ is a dimensionless constant. This gives $\beta =3/4$ (Eq. (3)). If one uses the Loitsyanskii integral $I_4$ instead, then one obtains $\beta =5/6$. \\

  The Birkhoff-Saffman and Loitsyanskii invariants were recently generalised for rotating, stratified and MHD turbulence \cite{dav1},\cite{dav2}. Therefore, the above mentioned values of the $\beta$ can be also valid for these types of anisotropic turbulence.  
 
  In this note we will consider a turbulence that emerges from distributed chaos with spontaneously broken space translational symmetry (homogeneity). First of all let us note that, unlike the inertial range, the asymptotic  Eq. (2) is under strong viscous influence. Therefore, in order to understand what dimensional parameter should be used in the Eq. (2) at the spontaneous breaking of the translational symmetry let us consider a final stage of decaying turbulence inside of a finite sphere of radius $R$ (with boundary conditions). One should not be confused by difference in asymptotic $k \rightarrow 0$ for the final decay case and the asymptotic Eq. (2). The main ideas related to the viscosity effects are similar in these two cases. For simplicity we will consider an incompressible fluid. In the final stage of decay only remnants of turbulence are remained and one can neglect the nonlinear convection terms in the Navier-Stoks equation for velocity field ${\bf u}  ({\bf x},t)$ \cite{my},\cite{saf} (but we preserve the pressure term because the long-range interactions in the {\it finite} volume can still exist due to the pressure) 
$$
\frac {\partial {\bf u}  ({\bf x},t)}{\partial t}= -\frac{1}{\rho} \nabla  p  ({\bf x},t) + \nu \nabla^2 {\bf u}  ({\bf x},t) \eqno{(6)}.
$$
Let us shift this equation by a constant vector ${\bf r}$
$$
\frac {\partial {\bf u}  ({\bf x} + {\bf r},t)}{\partial t}= -\frac{1}{\rho} \nabla  p  ({\bf x} + {\bf r},t) + \nu \nabla^2 {\bf u}  ({\bf x} + {\bf r},t) \eqno{(7)}
$$
 Let us multiply both sides of Eq. (6) by ${\bf u}  ({\bf x} + {\bf r},t)$  and both sides of Eq. (7) by ${\bf u}  ({\bf x},t)$ (the dot products), then make a summation of the two equations and then integrate the both sides over volume of the motion on the variable ${\bf x}$. If we denote the volume average of a function ${\bf A}  ({\bf x},t)$ as 
$$
\langle {\bf A}  ({\bf x},t) \rangle_{V} = \frac{\int_{V}  {\bf A}  ({\bf x},t) d{\bf x}}{\int_{V}  d{\bf x}} 
$$
where integration is over the volume of motion, then we obtain for the incompressible fluid
$$
\frac{d\langle {\bf u} ({\bf x},t) \cdot  {\bf u} ({\bf x} + {\bf r},t) \rangle_{V}}{d(\nu t)}=-2\langle {\boldsymbol \omega} ({\bf x},t) \cdot  {\boldsymbol \omega} ({\bf x} + {\bf r},t) \rangle_{V} \eqno{(8)}
$$  
where ${\boldsymbol \omega} ({\bf x},t)= \nabla \times {\bf u}  ({\bf x},t)$ and $\nu t$ is the difusive (viscous) time appropriate for consideration of quasi-invariants at the final stage of the turbulence decay (cf. Chapter 15 Ref. \cite{my}). Finally let us take integral $\int_{V} d{\bf r}$ on both sides of the equation (8)
$$
\frac{d\int_{V} \langle {\bf u} ({\bf x},t) \cdot  {\bf u} ({\bf x} + {\bf r},t) \rangle_{V} d{\bf r}}{d(\nu t)}=-2\gamma \eqno{(9)}
$$  
where
$$
\gamma = \int_{V} \langle {\boldsymbol \omega} ({\bf x},t) \cdot  {\boldsymbol \omega} ({\bf x} + {\bf r},t) \rangle_{V}  d{\bf r} \eqno{(10)}
$$
Of course, the average over volume eliminates dependence on ${\bf x}$ even in the case of the broken translational symmetry (cf. Introduction), but this average is equal to the ensemble average in the limit $R \rightarrow \infty $ only. If the Birkhof-Saffman integral is finite at this limit, then the right hand side of the Eq. (9) (the parameter $\gamma$) equals to zero \cite{bir},\cite{saf} and we have $$
\lim_{R\to\infty}\int_V \langle {\bf u} ({\bf x},t) \cdot  {\bf u} ({\bf x} + {\bf r},t) \rangle_{V}d{\bf r}=\int \langle {\bf u} ({\bf x},t) \cdot  {\bf u} ({\bf x} + {\bf r},t) \rangle d {\bf r}
$$
and
$$
\int \langle {\bf u} ({\bf x},t) \cdot  {\bf u} ({\bf x} + {\bf r},t) \rangle d {\bf r}=const
$$
at the final stage of decay of homogeneous isotropic turbulence. The same procedure can be applied to the field ${\boldsymbol \omega} ({\bf x},t)$ itself with conclusion that for $R \to \infty$ the parameter $\gamma$ is changing with the diffusion time $\nu t$ much slower than the integral $\int_{V} \langle {\bf u} ({\bf x},t) \cdot  {\bf u} ({\bf x} + {\bf r},t) \rangle_{V} d{\bf r}$.  All this indicates that the parameter $\gamma$ Eq. (10) can be considered as the parameter characterising spontaneous breaking of the space translational symmetry (let us recall that the Birkhof-Saffman invariant is associated with the conservation law of momentum and, through this, with the space translational symmetry).

   Then instead of 
the Eq. (5) we obtain from the dimensional consideration
$$
\upsilon (\kappa ) \simeq a_2~|\gamma|^{1/2}~\kappa^{1/2} \eqno{(11)}
$$
where $a_2$ is a dimensionless constant.
Equations (2) and (3) then give $\beta =1/2$. Observation of the stretched exponential power 
spectrum with the $\beta =1/2$ can be an indication of 
presence of the distributed chaos with the spontaneously broken translational symmetry (homogeneity).\\

 Where should one search for the distributed chaos with spontaneously broken space symmetries? First of all it can be the energy dissipation rate. Being a measure of velocity {\it gradients} the energy dissipation rate is naturally more sensitive to the distortions of the homogeneity than velocity field itself. One can also look to situation where the motion is bounded in one direction and free in the two other directions. This space configuration is rather appropriate for the spontaneous breaking of the space symmetries (a
 boundary layer or a channel flow, for instance, cf. Ref. \cite{b2}). The edges of magnetically confined plasmas also provide an interesting and practically important example of the shear flows. 
Another interesting case is the finite expanding universe.

\section{Isotropic turbulence}

\begin{figure}
\begin{center}
\includegraphics[width=8cm \vspace{-0.85cm}]{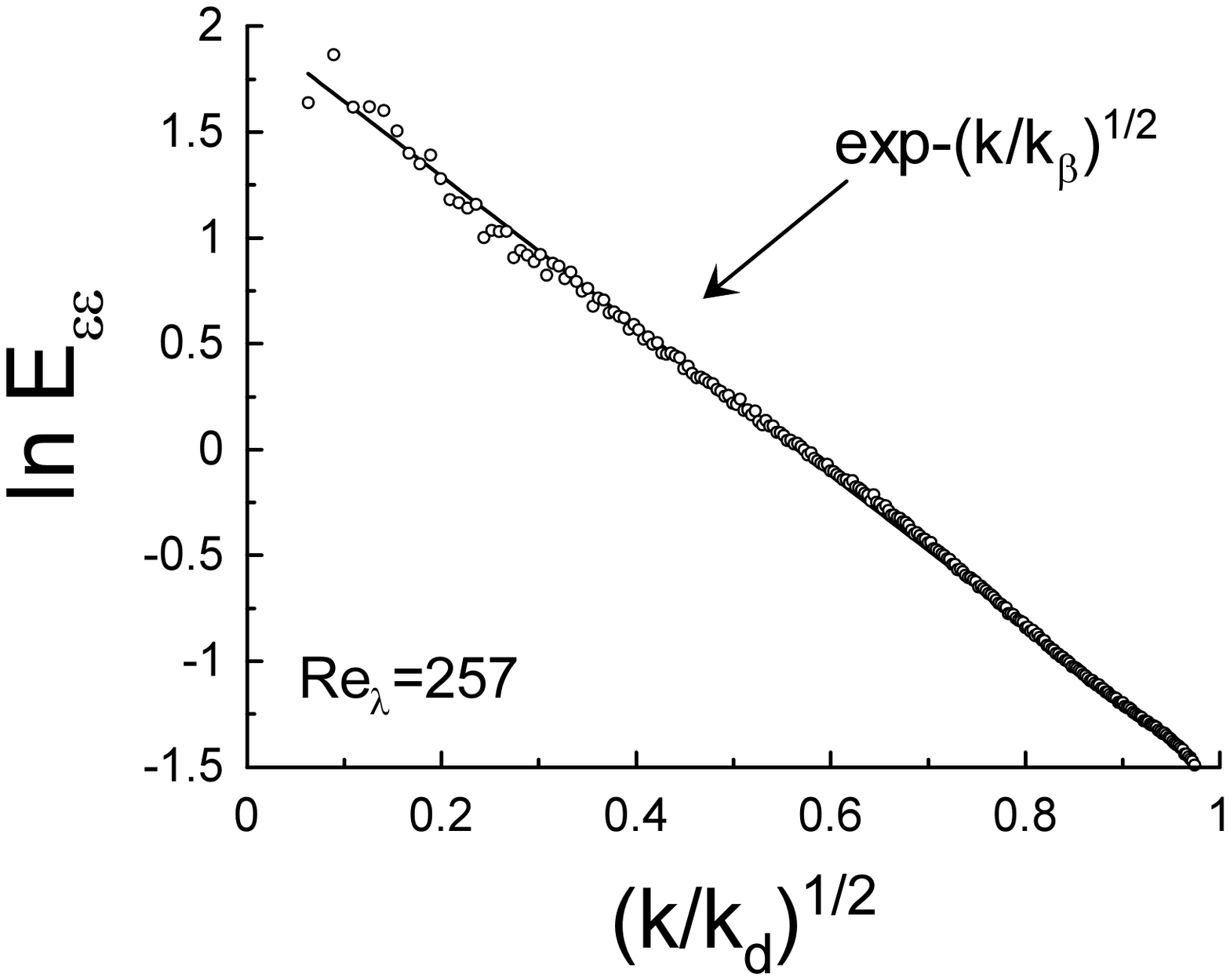}\vspace{-3.4cm}
\caption{\label{fig2} Logarithm of power spectrum of the energy dissipation rate field $\varepsilon$ at $Re_{\lambda}=257$ as function of $(k/k_d)^{1/2}$. The straight line indicates the exponential decay Eq. (1) with $\beta =1/2$. The data are taken from the same DNS \cite{ishi}.}. 
\end{center}
\end{figure}

In the direct numerical simulations of presumably homogeneous isotropic turbulence the motion is always approximately homogeneous. The energy dissipation rate $\varepsilon =2\nu e^2$, where 
$e^2=e_{ij}e_{ij}$ with $e_{ij}=(\partial u_i/\partial x_j+\partial u_j/\partial x_i )/2$, is more sensitive to inhomogeneities than velocity field. Therefore, the $\varepsilon$ field can exhibit the spontaneous breaking of the translational symmetry in situations where the velocity field itself is still sufficiently homogeneous. Figure 1 shows power spectrum of the energy dissipation rate field ($E_{\varepsilon \varepsilon}$) obtained in a DNS of a statistically stationary incompressible turbulence with periodic boundary conditions which can imitate homogeneous isotropic turbulence (see description of the DNS in Ref.  \cite{ishi}). The scales in this figure are chosen in order to represent the Eq. (1) as a straight line. The straight line in this figure indicates the Eq. (1) with $\beta =3/4$ (the Kolmogorov wavenumber $k_d = (\langle \varepsilon \rangle /\nu^3)^{1/4}$, and $k_{\beta} \simeq 0.234k_d$). Fig. 1 shows the data obtained at Reynolds number $Re_{\lambda } =167$. One can see that at this value of $Re_{\lambda }$ the $\varepsilon$ field exhibits good homogeneity. Figure 2 shows the data of the same DNS at 
$Re_{\lambda } =257$. The straight line in this figure corresponds to the Eq. (1) with $\beta =1/2$ ($k_{\beta} \simeq 0.079k_d$) that indicates the spontaneous breaking of the translational symmetry in the $\varepsilon$ field. 

The energy dissipation rate plays crucial role in development of inertial range \cite{my} near the range of the distributed chaos for high Reynolds numbers \cite{b1}. Therefore, this spontaneous breaking of the translational symmetry (homogeneity) in the $\varepsilon$ field can have very significant consequences for the inertial range (especially taking into account that the wavenumbers of the inertial range are smaller than those of the distributed chaos range and, therefore, are more vulnerable to the homogeneity distortions). It seems that not only local isotropy \cite{my}, \cite{ps} but also local homogeneity should be the matter of concern for the inertial range (and may be more significant one, cf. Introduction). \\

\begin{figure}
\begin{center}
\includegraphics[width=8cm \vspace{-0.63cm}]{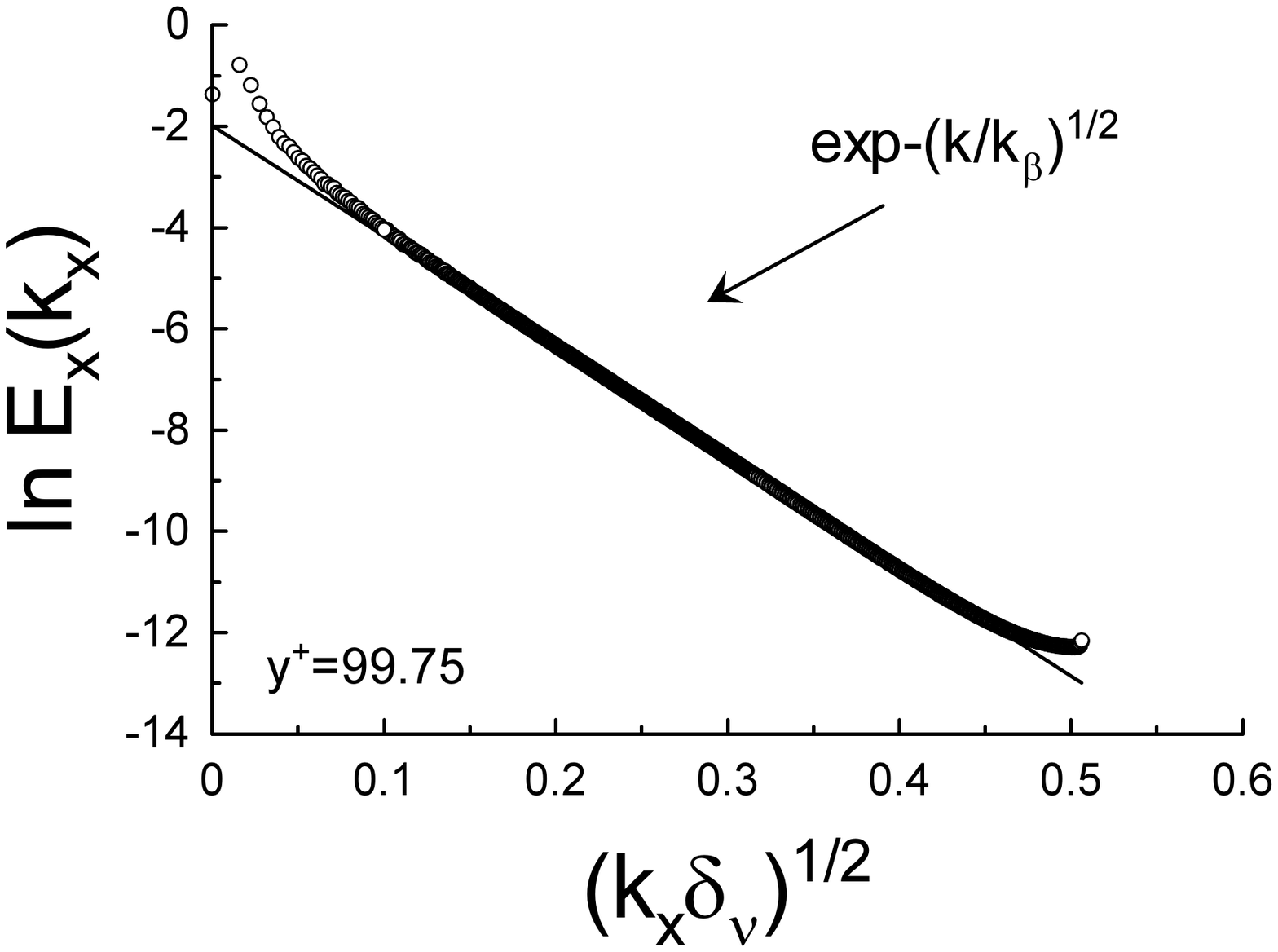}\vspace{-3.8cm}
\caption{\label{fig3} Logarithm of the streamwise power spectrum of streamwise component of the velocity at $y^{+} =99.75$ as function of $(k_{x}\delta_{\nu})^{1/2}$. The straight line indicates the exponential decay Eq. (1) with $\beta =1/2$.}
\end{center}
\end{figure}
 
\begin{figure}
\begin{center}
\includegraphics[width=8cm \vspace{-0.82cm}]{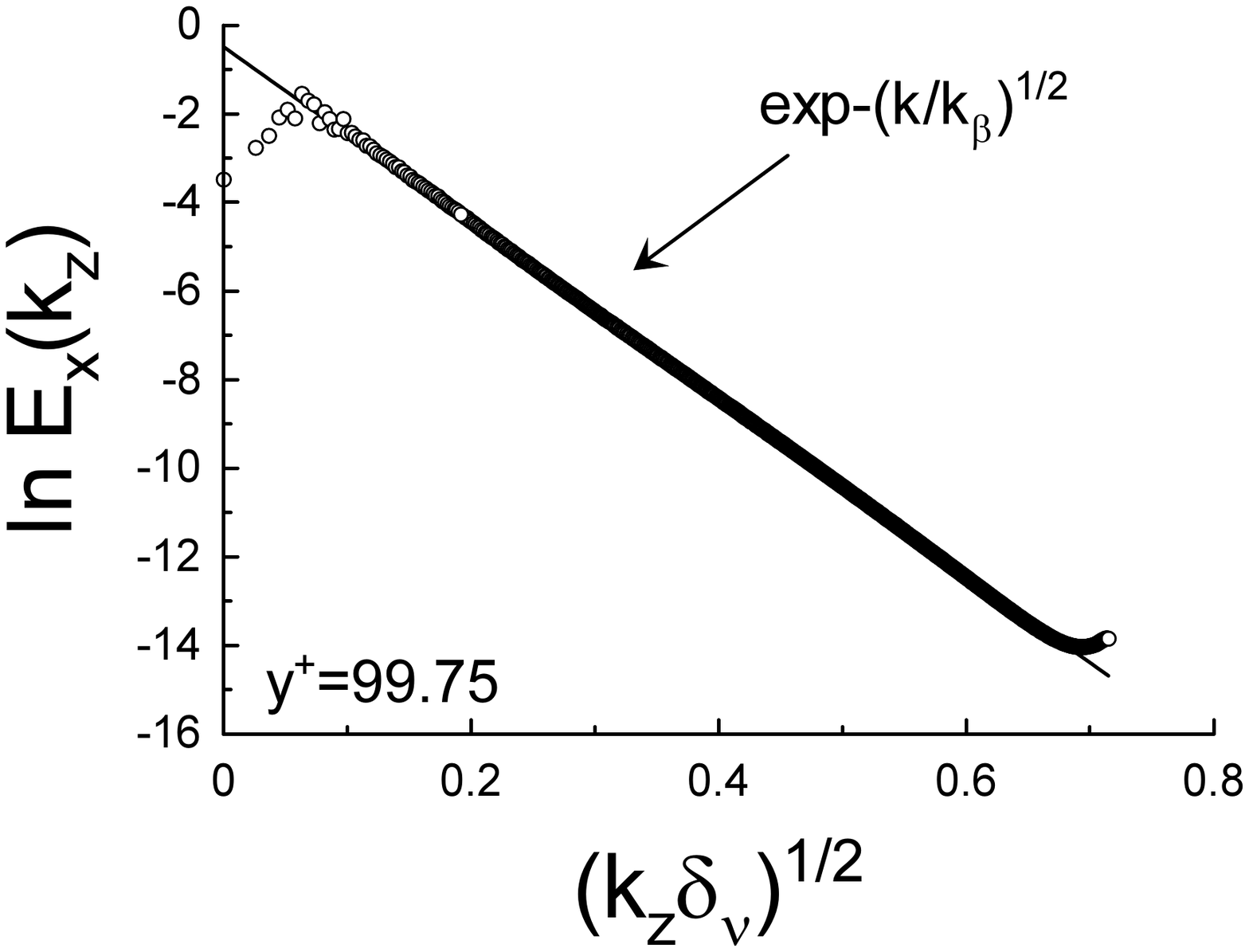}\vspace{-4cm}
\caption{\label{fig4} Logarithm of the spanwise power spectrum of the streamwise component of the velocity at $y^{+} =99.75$ as function of $(k_{z}\delta_{\nu})^{1/2}$.}. 
\end{center}
\end{figure}

\begin{figure}
\begin{center}
\includegraphics[width=8cm \vspace{-0.95cm}]{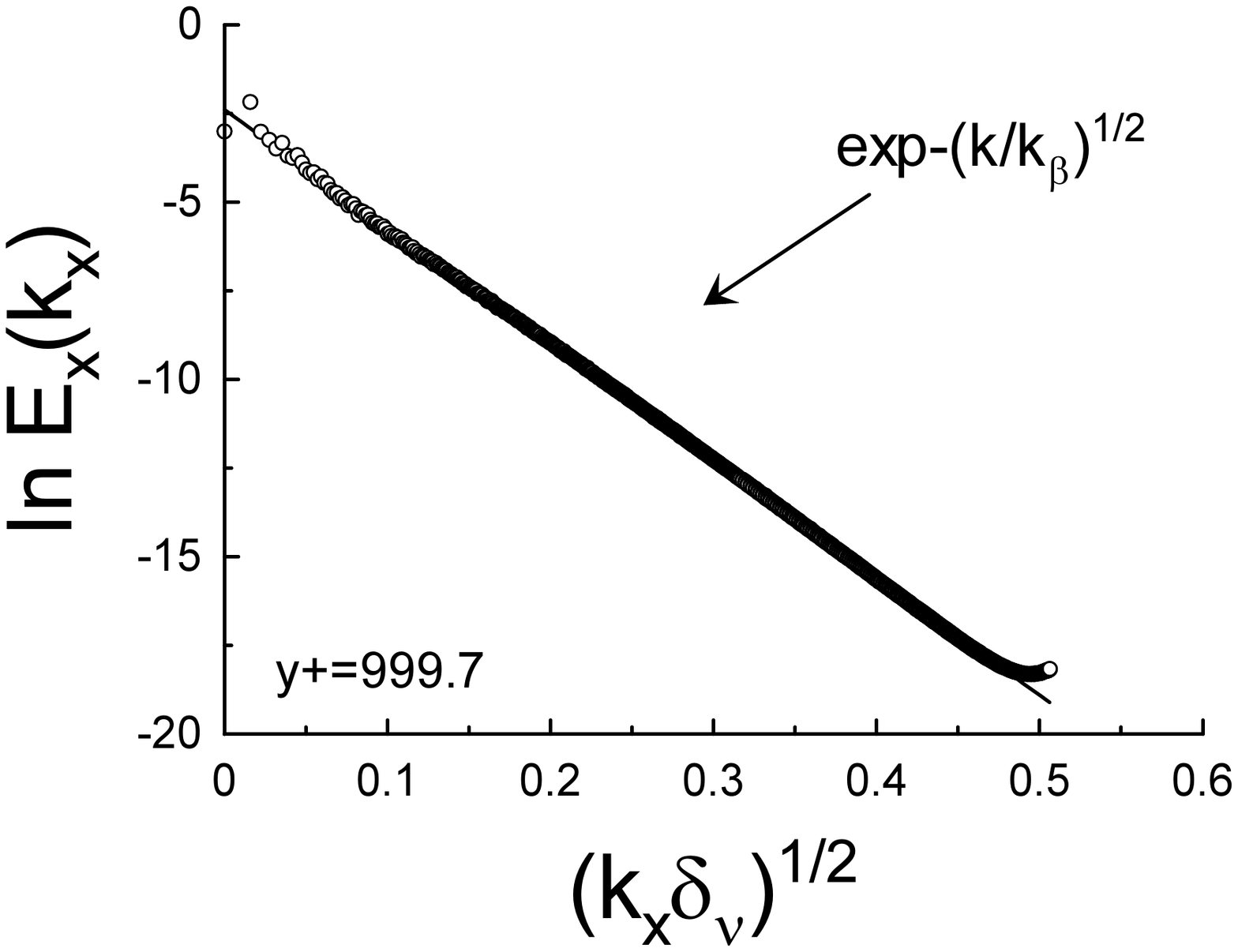}\vspace{-4cm}
\caption{\label{fig5} The same as in Fig. 3 but for $y^{+} =999.7$.}
\end{center}
 \end{figure}
 
 \begin{figure}
\begin{center}
\includegraphics[width=8cm \vspace{-1.5cm}]{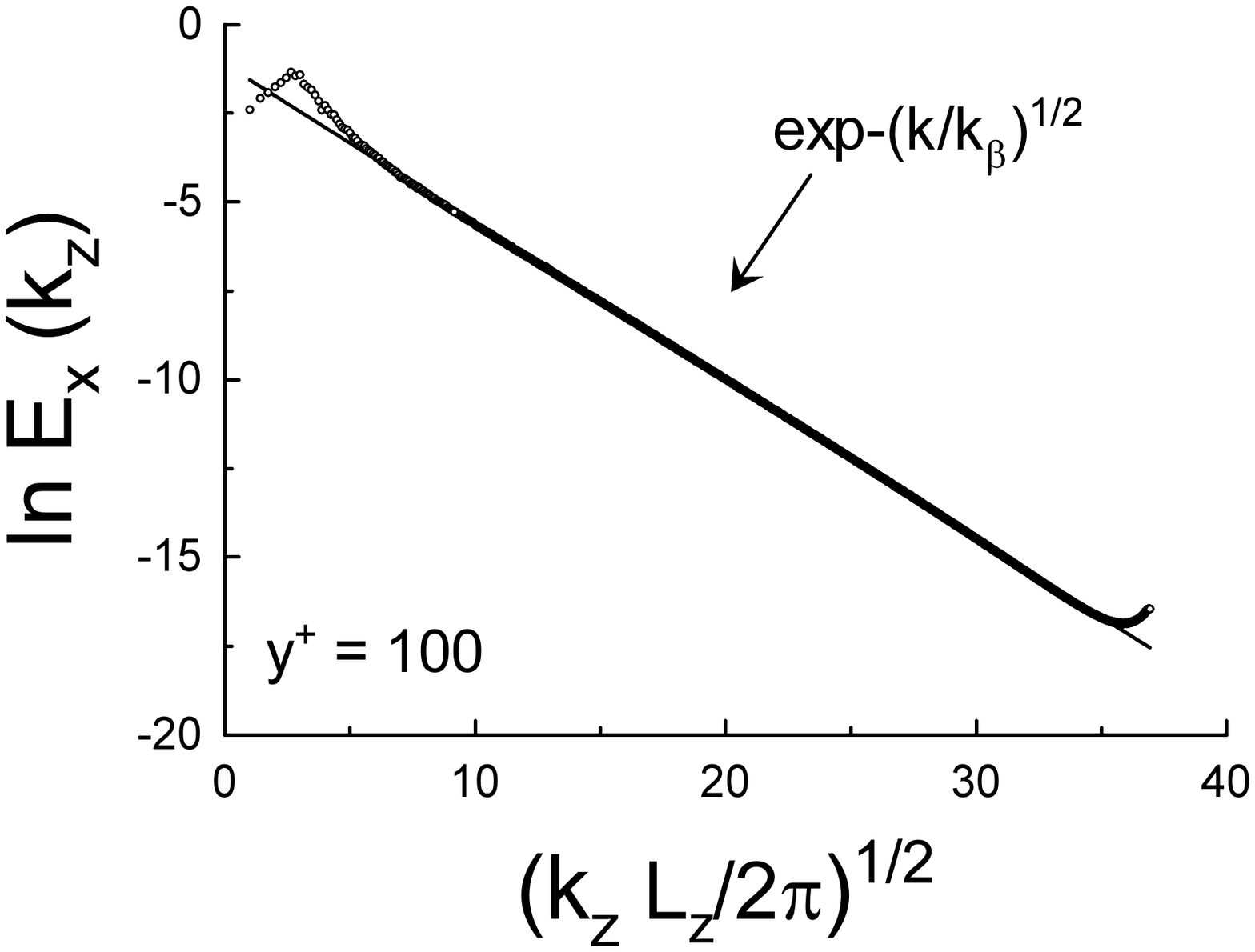}\vspace{-3cm}
\caption{\label{fig6} The same as in Fig. 4 but for direct numerical simulations of fully developed boundary layer flow $Re_{\tau}\simeq 1990$ (  $Re_{\theta}\simeq 6499$, $L_z$ is the DNS box dimension along the $z$ axis). The data were taken from the site \cite{tor}. 
}
\end{center}
\end{figure}

\begin{figure}
\begin{center}
\includegraphics[width=8cm \vspace{-0.8cm}]{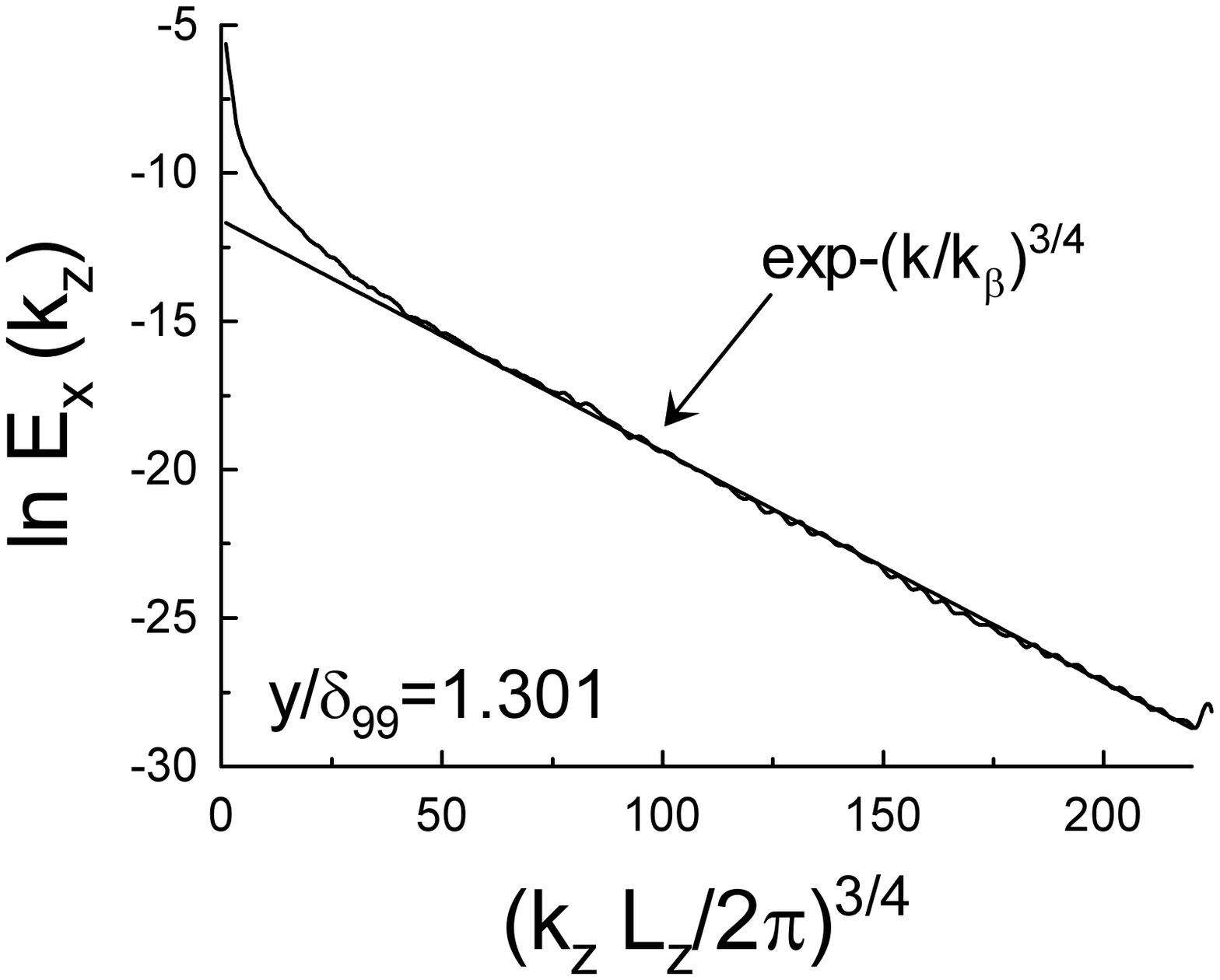}\vspace{-4cm}
\caption{\label{fig7} The same as in Fig. 6 but outside the boundary layer (at $y/\delta_{99}=1.301$). The straight line indicates the homogeneous value of $\beta =3/4$. The data were taken from the same site \cite{tor}. 
}
\end{center}
\end{figure}
\begin{figure}
\begin{center}
\includegraphics[width=8cm \vspace{-0.99cm}]{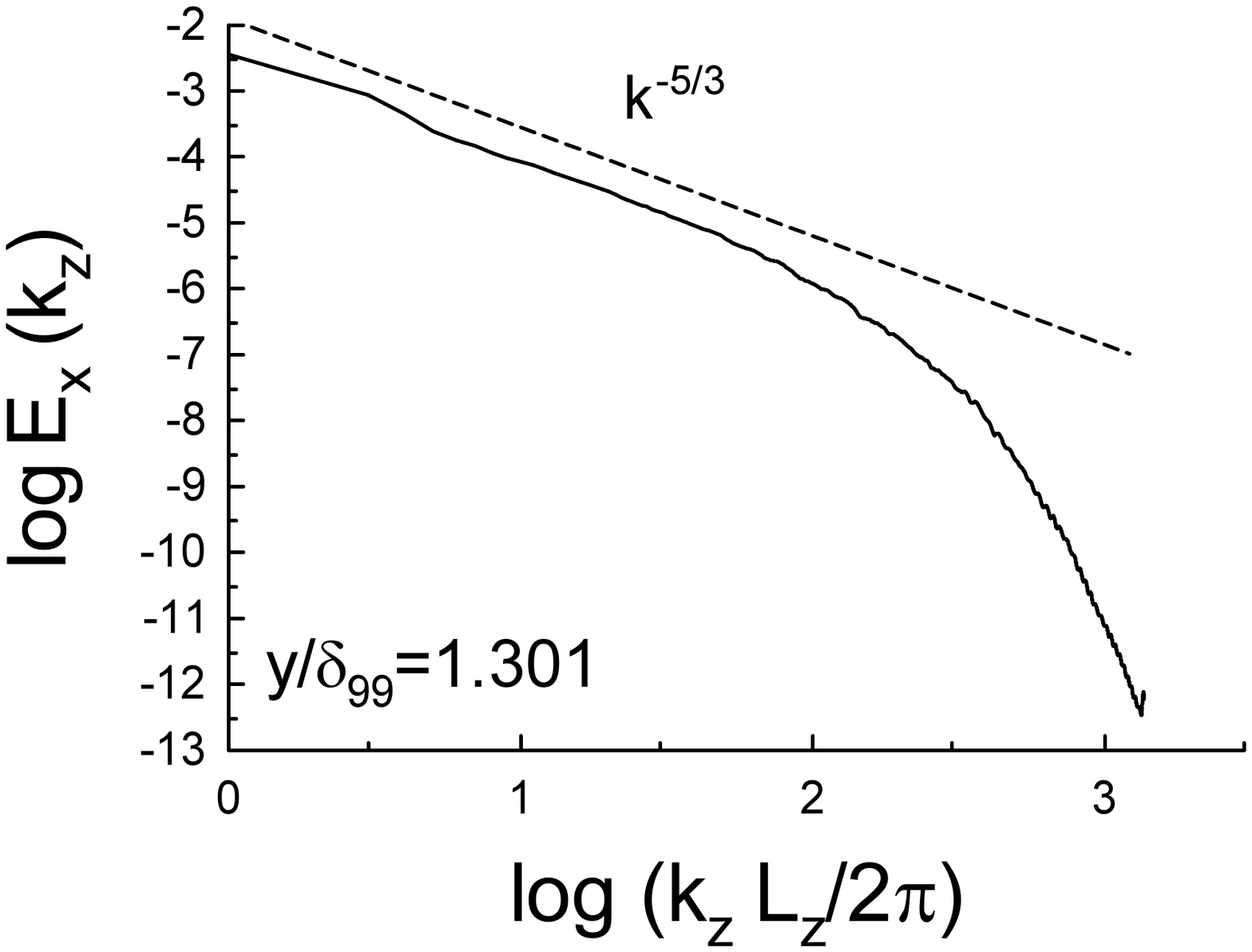}\vspace{-3.8cm}
\caption{\label{fig8} The same as in Fig. 7 but in the log-log scales. The dashed straight line indicates the homogeneous power-law $E(k) \propto k^{-5/3}$.  
}
\end{center}
\end{figure}
\section{Channel and boundary layer flows}
 
 In figures 3-5 we will use the data (taken from the site \cite{cf}) of a direct numerical simulation (DNS) of walls (planes) bounded turbulent flow with the friction velocity Reynolds number $Re_{\tau} \sim 1000$ (channel flow, see also Refs. \cite{men1},\cite{men2},\cite{kmm}), and the data (taken from the site \cite{tor}) of a direct numerical simulation of fully developed boundary layer flow $Re_{\tau}\simeq 1990$ $Re_{\theta}\simeq 6499$ (see also Refs. \cite{sim},\cite{bsj},\cite{sjm}). 
 
 The power spectra shown in the Figures 3-5 correspond to the distances $y^{+} = 99.75$ and $y^{+} = 999.7$ from a wall of the channel flow (belonging to the so-called 'logarithmic law' range, i.e. far away from the near wall range with its complex structures) \cite{cf}. The power spectrum shown in Fig. 6 corresponds to distance $y^{+} \simeq 100$ (the fully developed boundary layer flow \cite{tor}).  
 
 Figure 3 shows streamwise ($k_x$) power spectrum of the streamwise component of the velocity.     The scales in this figure are chosen in order to represent the Eq. (1) with $\beta =1/2$ as a straight line. The $ \delta_{\nu} = \nu/u_{\tau}$ is the viscous length scale, $u_{\tau}$ is the friction velocity. Figure 4 shows spanwise ($k_z$) power spectrum of the same streamwise component of the velocity. The distance from the wall is $y^{+} =99.75$ (in viscous units) for the both cases. $k_{\beta} \simeq 0.0021 \delta_{\nu}^{-1}$ for the streamwise ($k_x$) power spectrum (Fig. 3), and $k_{\beta} \simeq 0.0025 \delta_{\nu}^{-1}$ for spanwise ($k_z$) power spectrum (Fig. 4). For the streamwise ($k_x$) power spectrum this type of spectrum is observed also at the end of the 'logarithmic law' range: $y^{+} =999.7$, as one can see at Fig. 5. The parameter $k_{\beta} \simeq 0.0012 \delta_{\nu}^{-1}$ for the streamwise ($k_x$) power spectrum at the distance $y^{+} =999.7$.

   For comparison, the figures 7 and 8 show data from the same DNS as Fig. 6  but outside the boundary layer at $y/\delta_{99}=1.301$ (where $\delta_{99}$ is the thickness of the velocity boundary layer). For the ln-linear scales in the Figure 7 the straight line indicates the exponential decay Eq. (1) with $\beta =3/4$. Figure 8 shows the same spectrum but now in the log-log scales. In the Fig. 8 the straight line indicates the Kolmogorov's power-law spectrum  $k^{-5/3}$.  In the terms of the Ref. \cite{b1} the picture shown in the Figs. 7 and 8 is typical for homogeneous turbulence: coexistence of the range of the distributed chaos dominated by the Birkhoff-Saffman integral (characterized by $\beta=3/4$, Fig. 7) and of the inertial (Kolmogorov) range (Fig. 8). This is precisely what one can expect outside the boundary layer.

 The exponent $\beta =1/2$ provides rather good agreement with the data shown in the Figs. 3-6, that can be considered as an indication of strong presence of the distributed chaos with spontaneously broken translational space symmetry (homogeneity). The value of $k_{\beta}$ seems to be non universal in this case (cf. with the isotropic homogeneous case \cite{b1}).

\section{The edges of magnetically confined plasmas} 

Interesting and practically important example of the shear flows is now under vigorous experimental investigation at the edges of magnetically confined plasmas in different fusion devices. Chaotic motion at the edges degrades performance of these devices. 
   Measurements of ion saturation current is used in order to obtain dynamic information there. It was discovered that frequency spectra of the ion saturation current fluctuations are collapsed in a singular functional shape for the toroidal devices (stellarators and tokamaks) \cite{ped}. The frequency spectra obtained by a probe with a fixed space location in a shear flow reflex the spatial spectra of the structures moving near the probe (the Taylor hypothesis \cite{my}). Therefore, the frequency stretched exponential spectrum
$$
E(f) \propto \exp-(f/f_{\beta})^{1/2}  \eqno{(12)}   
$$
reflex the wavenumber spectrum Eq. (1) with $\beta = 1/2$. In this case the differences in the spectra between the different devices can be related to the difference in the parameter $f_{\beta}$ only (after normalization of the amplitudes). 

  Figure 9 shows the normalized and frequency rescaled power spectra of ion saturation current. The measurements were made at the plasma edges of several fusion devices (see the legend). The authors of the Ref. \cite{ped} (where these data came from) rescaled the frequency with the parameter $\lambda$. The cumulative data for different devices (used in the Fig. 9) were taken from the Ref. \cite{mm1}. The scales in Fig. 9 are chosen in order to represent the Eq. (12) as a straight line.

\section{Large-scale distribution of the galaxies}

It is an old and rather natural idea that turbulence played a crucial role in formation of galaxies and clusters of galaxies \cite{w}\cite{g} (see for more recent developments Refs. \cite{gs},\cite{z},\cite{psb} and references therein). In the finite expanding universe one can expect the spontaneous breaking of the space translational symmetry (homogeneity) at certain stage. To check role of the distributed chaos one can use the power spectrum of a distribution of galaxies. This spectrum is the Fourier transform of the two-point correlation function $\xi ({\bf r})$:
$$
P(\bf{k})= \frac{n}{(2\pi)^{3/2}} \int \xi ({\bf r}) e^{i{\bf k}{\bf r}} d {\bf r}  \eqno{(13)}
$$
with $n$ representing the average density and the correlation function $ \xi ({\bf r})$ is defined by equation
$$
dp=n(1+\xi (r))dV  \eqno{(14)}
$$
where $dp$ is probability of finding a galaxy in a volume $dV$ separated
by a distance $r$ (isotropic case) from a given one (see Ref. \cite{m}, for instance).

\begin{figure} 
\begin{center}
\includegraphics[width=8cm \vspace{-0.6cm}]{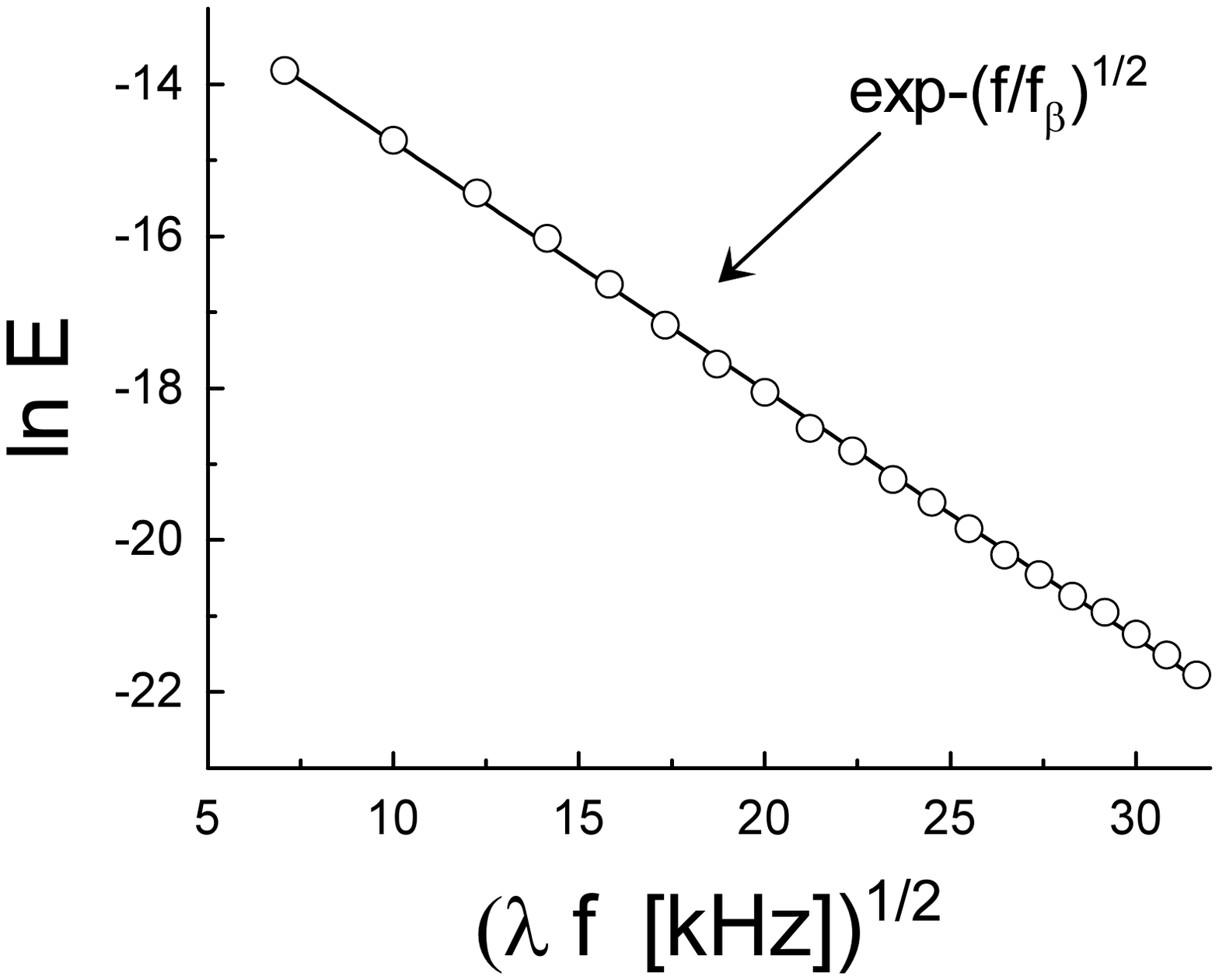}\vspace{-4.4cm}
\caption{\label{fig9} Power spectrum of ion saturation current. The fusion devices - JET: $\lambda =4.5$,  $W7-AS^{(1)}: \lambda = 3.5$, $W7-AS^{(2)}: \lambda = 4.5$, TJ-I: $\lambda =1$, TJ-IU: $\lambda =3$,  The straight line is drawn in order to indicate a stretched exponential decay Eq. (12).} 
\end{center}
\end{figure}

\begin{figure} 
\begin{center}
\includegraphics[width=8cm \vspace{-0.75cm}]{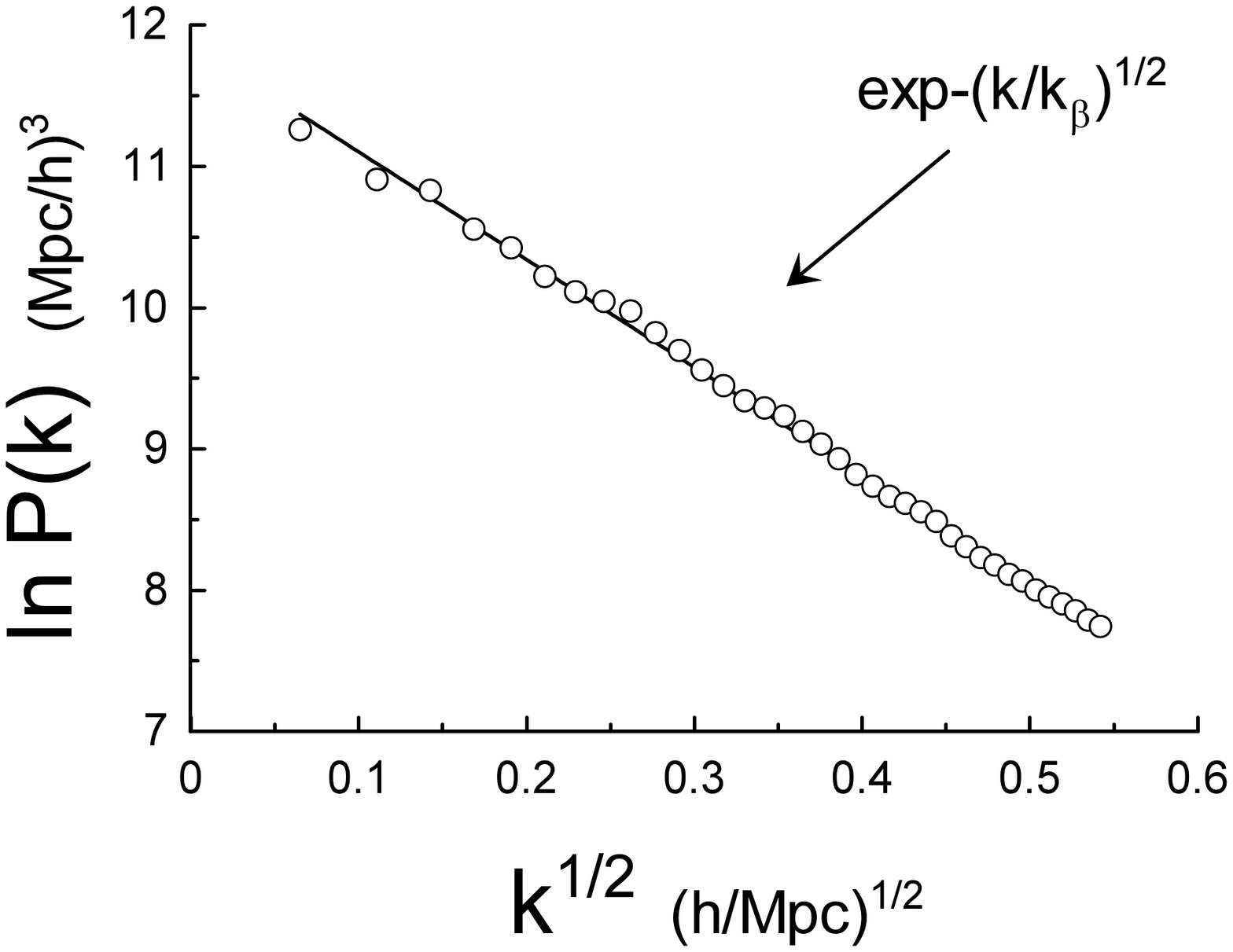}\vspace{-4cm}
\caption{\label{fig10} The power spectrum $P(k)$ for the galaxies distribution SDSS-III BOSS DR11 (post reconstruction data were taken from \cite{ander}). The straight line indicates the exponential decay Eq. (1) with $\beta =1/2$.} 
\end{center}
\end{figure}
 In recent paper Ref. \cite{ander} the $P(k)$ spectrum was calculated for the data of the BOSS -  Baryon Oscillation Spectroscopic Survey. This is the largest component of the Sloan Digital Sky Survey SDSS-III (DR11 sample of the BOSS $\sim$ one million galaxies, redshift range $0.2 < z < 0.7$). 
 This spectrum is shown in Fig. 10. The post reconstruction data were taken from \cite{ander} (available at site https://www.sdss3.org/science, $\to$ BOSS $\to$ Anderson-2013-CMASSDR11-power-spectrum-post-recon-1.dat). The straight line in the Fig. 10 indicates the exponential decay Eq. (1) with $\beta =1/2$ ($k_{\beta} \simeq 0.017~ h/Mpc$). 
 
 Modern cosmology has a fundamental problem. For relatively small scales the power spectrum $P(k)$ exhibits scaling (self-similarity) whereas on large scales there is no scaling \cite{m}. Is there a law describing $P(k)$ on the large scales?  One can see that the turbulent distributed chaos with spontaneously broken space translational symmetry provides a quantitative answer to this question. Just the sufficiently large scales 'feel' the finite size of the universe (the boundaries), that results in the spontaneous breaking of the translational space symmetry (homogeneity) and in the nonscaling spectrum $P(k)$ of the type Eq. (1) with $\beta=1/2$.

\section{Acknowledgement}

I thank T. Ishihara, D. J. Graham, M. Lee, N. Malaya, R.D. Moser, G. Eyink, C. Meneveau, M. P. Simens J. Jimenez, S. Hoyas, Y. Mizuno, G. Borrell, and J. A. Sillero for sharing their data. I also thank P. A. Davidson, C. Meneveau, J. Schumacher and K. R. Sreenivasan for comments and encouragement.

\end{document}